\documentclass[useAMS,usenatbib]{mn2e}
\usepackage{graphicx}
\usepackage{epsfig}

\title[{\it Suzaku} observation of G272.2$-$3.2]{{\it Suzaku} study of centrally brightened supernova remnant G272.2$-$3.2}
\author[A. Sezer, F.~G\"{o}k]{A. Sezer,$^{1,2}$\thanks{E-mail: aytap.sezer@uzay.tubitak.gov.tr (AS); gok@akdeniz.edu.tr
(FG)} F.~G\"{o}k $^{3}$
\footnotemark[1]\thanks{This file has been amended to highlight
the proper use of \LaTeXe\ code with the class file. These changes
are for illustrative purposes and do not reflect the
original paper by A.~Sezer.}\\
$^{1}$T\"UB\.ITAK Space Technologies Research Institute, ODTU
Campus, Ankara, 06531, Turkey\\
$^{2}$Bo\~gazi\d{c}i University, Faculty of Art and Sciences,
Department of Physics, \.Istanbul, 34342, Turkey\\
$^{3}$Akdeniz University, Faculty of Sciences, Department
of Physics, Antalya, 07058, Turkey\\
}

\begin{document}

\date{}

\pagerange{\pageref{firstpage}--\pageref{lastpage}} \pubyear{2011}

\maketitle

\label{firstpage}

\begin{abstract}
In this work, the results from {\it Suzaku} observation of
Galactic supernova remnant G272.2$-$3.2 are presented. Spectra of
G272.2$-$3.2 are well fitted by a single-temperature variable
abundances non-equilibrium ionization (VNEI) model with an
electron temperature $kT_{\rm e}\sim0.77$ keV, ionization
timescale $\tau\sim6.5\times10^{10}$ ${\rm cm^{-3}}$s and
absorbing column density $N_{\rm H}\sim1.1\times10^{22}$
cm$^{-2}$. We have detected enhanced abundances of Si, S, Ca, Fe
and Ni in the center region indicating that the X-ray emission has
ejecta origin. We estimated the electron density $n_{\rm e}$ to be
$\sim$0.48$f^{-1/2}$ $\rm cm^{-3}$, age $\sim$$4300f^{1/2}$ yr and
the X-ray total mass $M_{\rm x}=475f^{1/2}$ {M\sun} by taking the
distance to be d=10 kpc. To understand the origin of the
centrally-peaked X-ray emission of the remnant, we studied radial
variations of the electron temperature and surface brightness. The
relative abundances in the center region suggest that G272.2$-$3.2
is the result of a Type Ia supernova explosion.

\end{abstract}

\begin{keywords}
ISM: supernova remnants$-$ISM:individual G272.2$-$3.2$-$X-rays:ISM
\end{keywords}

\section{Introduction}

The Galactic supernova remnant (SNR) G272.2$-$3.2 was discovered
in X-rays with {\it ROSAT} All-Sky Survey \citep{b5}, it has
centrally filled X-ray morphology and a thermally dominated X-ray
spectrum. \citet{b6} found that the electron temperature was
between 1.0$-$1.5 keV from the {\it ROSAT} PSPC data. \citet{b10}
confirmed that the nature of the emission was shock-heated and the
nebulosity was an SNR by measuring the [S\,{\sc ii}]/H${\alpha}$
ratio and detecting emissions from [N\,{\sc ii}] 658.3 nm and
[O\,{\sc ii}] 732.5 nm in the optical band. \citet{b12} carried
out a series of radio observations with a Parkes radio telescope,
an Australia Telescope Compact Array ({\it ATCA}) and a Molonglo
Observatory Synthesis Telescope ({\it MOST}). From these
observations they show that the radio spectral index of this
remnant was typical of shell-type SNRs, $\alpha\sim0.55$, and
almost circular in appearance with a diameter of $\sim$15 arcmin.
The remnant consists of faint filaments and patches of emission
with a low surface brightness as well as bright blobs that
correlate well with the brightest optical filaments and of a
diffuse emission component produced by shock-accelerated
electrons. \citet{b4} found that the X-ray emission from
G272.2$-$3.2 could be best described by a non-equilibrium
ionization (NEI) model with an electron temperature of about 0.7
keV, an ionization timescale of 3200 cm$^{-3}$ yr and a relatively
high column density, $N_{\rm H}\sim10^{22}$ atoms$^{-2}$, from
{\it ASCA} and {\it ROSAT} observations. They discussed cloud
evaporation and thermal conduction models to explain the centrally
peaked X-ray morphology of the remnant. Later, from an X-ray study
of G272.2$-$3.2 with {\it Chandra}, \citet{b13} reported that the
X-ray spectra of the outer shell regions showed normal
compositions being consistent with the shocked interstellar
medium, while central emission showed elevated abundances
suggesting reverse shocked stellar ejecta from Type Ia supernova
(SN).  \citet{b22}, based on {\it Chandra}'s observation,
confirmed that G272.2$-$3.2 is a Type Ia origin.

The distance to G272.2$-$3.2 is not well known. It was estimated
to be $1.8^{+1.4}_{-0.8}$ kpc from observed interstellar
absorption by \citet{b6}. From the statistical analysis,
\citet{b4} calculated 2 kpc, which is in agreement with the
distance found by \citet{b6}. They also obtained an upper limit of
10 kpc using optical color excess with a distance of roughly 0.2
mag kpc$^{-1}$ in the direction of G272.2$-$3.2 and adopted
intermediate distance of 5 kpc to this remnant.

In this study, we investigate the nature of X-ray emission of
G272.2$-$3.2 which is characterized by an apparent centrally
brightened X-ray morphology and thermally dominated X-ray emission
by utilizing the superior spectral capabilities for diffuse
sources of X-ray Imaging Spectrometers (XIS: \citet{b8}) onboard
{\it Suzaku} satellite \citep{b9}. The structure of the paper is
as follows; in Section 2, we describe the {\it Suzaku} observation
and data reduction. Image and spectral analysis are presented in
Section 3 and 4, respectively. In Section 5, we discuss the
physical properties of the thermal X-ray emitting plasma (5.1),
possible reasons of centrally peaked morphology (5.2) and finally,
relative abundances in the ejecta (5.3).

\section[]{Observation and Data reduction}
{\it Suzaku} observed G272.2$-$3.2 on 2011 May 28 by the XIS. The
observation ID and exposure time are 506060010 and 130 ksec,
respectively. The XIS has four CCDs: three of them (XIS0, 2, and
3) are front-illuminated (FI) and one (XIS1) is back-illuminated
(BI). The XISs are sensitive to the 0.2$-$12.0 keV energy band
with a 17.8$\times$17.8 arcmin$^{2}$ field of view (FOV). In
November of 2006, XIS2 was damaged and taken off-line, therefore
data taken after the 2007 observation were taken with only the
remaining three XISs. The XIS was operated in the normal
full-frame clocking mode. Two corners of each XIS CCD have an
$^{55}$Fe calibration source which can be used to calibrate the
gain and test the spectral resolution of data taken using this
instrument.

Reduction and analysis of the data were performed by following the
standard procedure using the {\sc headas} v6.4 software package,
and spectral fitting was performed with {\sc xspec} v.11.3.2
(Arnaud 1996). All of the data were reprocessed, referring to the
CALDB as of July 9, 2008. The redistribution matrix files (RMFs)
of the XIS were produced by {\sc xisrmfgen}, and auxillary
response files (ARFs) by {\sc xissimarfgen} \citep{b7}.

\section{Image Analysis}
Fig. 1 shows an XIS1 image of G272.2$-$3.2 in the 0.3$-$10 keV
energy band. From this figure, we see brighter emission in the
central region (within $\sim$3.8 arcmin radius) and a relatively
fainter emission in the outer part. Central and outer regions are
shown by solid black circles centered at
$\rmn{RA}(2000)=09^{\rmn{h}} 06^{\rmn{m}} 47^{\rmn{s}}$,
$\rmn{Dec.}~(2000)=-52\degr 06\arcmin 05\arcsec$. Dashed white
circles with sizes of 0$-$1.5, 1.5$-$2.5, 2.5$-$3.5, 3.5$-$4.5,
4.5$-$5.5 arcmin are chosen to obtain the radial variations of the
electron temperature $kT_{\rm e}$ and surface brightness. The
black dashed circle with a radius of 1.5 arcmin represents the
background region, $\rmn{RA}(2000)=09^{\rmn{h}} 07^{\rmn{m}}
18^{\rmn{s}}$, $\rmn{Dec.}~(2000)=-52\degr 14\arcmin 27\arcsec$,
used for spectral analysis. The black dashed square indicates the
FOV of the XIS1.

\section{Spectral Analysis}

The spectrum is extracted first from all over the remnant
(hereafter whole region) with a radius of 7.3 arcmin, then from
the central region with the brightest X-rays and finally from the
outer region where the X-rays are fainter, this is indicated by
solid black circles (see Fig.1) for each of the XISs. The spectra
are grouped with a minimum of 120 counts bin$^{-1}$ for the whole
region and 50 counts bin$^{-1}$ for the central and outer regions.

For the whole region, we applied the VNEI model, a model in {\sc
xspec} for a NEI collisional plasma with variable abundances
\citep{b3}, modified by interstellar absorption using cross
sections from \citet{b14} in the 0.3$-$10 keV energy range. The
absorption column density $N_{\rm H}$, electron temperature
$kT_{\rm e}$, ionization timescale $\tau=n_{\rm e}t$, where
$n_{\rm e}$ is the electron density and $t$ is the elapsed time
after the plasma was heated up, and normalization were set as free
parameters and all elemental abundances were fixed to their solar
values of \citet{b1}. From this fit we obtained a reduced
$\chi^{2}$ of 1.81 for 752 degrees of freedom (dof). Then, we
allowed O, Ne, Mg, Si, S and Fe abundances to vary, since these
abundances have appeared to differ from their solar values and the
line features were evident in the spectra, while other elemental
abundances were frozen to their solar values. In this case, the
spectral fit has significantly improved with reduced $\chi^{2}$ of
1.02 for 746 dof. We repeated the same steps for the central and
outer regions. The parameter values obtained for each region are
listed in Table 1, and the errors quoted are 90 per cent
confidence limits. The background-subtracted FI XIS (XIS0 and
XIS3) spectra of each region in 0.3$-$10 keV are shown in Fig. 2.

Radial variations of the electron temperature and surface
brightness are plotted in Fig. 3. During spectral fittings, to get
an estimate of possible temperature variation across the SNR, we
fixed the absorbing column density and the ionization timescale to
the values of the whole region.

\begin{table*}
\centering
 \begin{minipage}{140mm}
  \caption{Best-fitting parameters of the spectral fitting in the 0.3$-$10 keV
energy band for all over the remnant (whole), its centre and outer
regions with an absorbed VNEI model with corresponding errors at
90 per cent confidence level (2.7 $\sigma$).}
 \begin{tabular}{@{}ccccc@{}}
  \hline
      Parameters & Whole& Centre & Outer & \\
 \hline
 $N_{\rm H}$($\times10^{22}$$\rm cm^{-2})$ &1.07$\pm 0.02$&1.14$\pm 0.02$ &0.96$\pm 0.02$ & \\
$kT_{\rm e}$(keV) & 0.77 $\pm 0.02$& 0.83 $\pm 0.03$&0.76 $\pm 0.02$ &\\
 O (solar) & 1.4 $\pm 0.4$& (1)& 0.6 $\pm 0.2$&\\
 Ne (solar) & 0.6 $\pm 0.1$&0.2 $\pm 0.1$ &0.4 $\pm 0.1$ &\\
 Mg (solar) & 0.7 $\pm 0.1$&0.7 $\pm 0.1$ & 0.6 $\pm 0.1$&\\
 Si (solar)& 1.3 $\pm 0.1$& 2.0 $\pm 0.1$& 0.8 $\pm 0.1$&\\
 S (solar)& 2.2 $\pm 0.2$& 4.0 $\pm 0.2$&1.2 $\pm 0.1$ &\\
 Ca (solar) & (1)&1.8 $\pm 1.1$ & (1)&\\
 Fe (solar)& 1.3 $\pm 0.2$& 1.96 $\pm 0.11$& 0.8 $\pm 0.1$&\\
 Ni (solar)& (1)& 3.9 $\pm 0.9$&(1)&\\
$n_{\rm e}t$($\times10^{10}$$\rm cm^{-3}s)$& 6.5 $\pm 0.6$&5.3$\pm 0.5$ &6.2$\pm 0.6$ &\\
normalization& 0.19$\pm 0.02 $& 0.17 $\pm 0.01 $& 0.16$\pm 0.01$ & \\
Flux\footnote{{ Flux corrected for Galactic absorption} in the $0.3-10$ keV energy band in the unit of $10^{-11}$ erg $\rm s^{-1}$$\rm cm^{-2}$.}& 6.8$\pm 0.1$ &  7.7$\pm 0.1$&  4.3$\pm 0.1$&\\
 $\chi^{2}$/dof  &760.7/746=1.02 &917.2/686=1.34 & 733.7/945=0.78& \\
\hline
\end{tabular}
\end{minipage}
\end{table*}

\section{Discussion and Conclusions}
In this paper, we report the results of high quality X-ray spectra
and detailed analysis of G272.2$-$3.2 using {\it Suzaku} XIS
observation. The X-ray spectra can be represented with a
non-equilibrium ionization plasma (VNEI) model with an electron
temperature of $kT_{\rm e}\sim0.77$ keV, high absorbing column
density and a relatively small ionization timescale, less than
$10^{12}$ ${\rm cm^{-3}}$s. We found clear K-shell lines of O, Ne,
Mg, Si, S, Ca, Fe and Ni in the 0.3$-$10 keV band spectra, as
shown in Fig. 2. The central region is enhanced in Si, S, Ca, Fe
and Ni, as shown in Table 1. This fact suggests that the X-ray
emission originating from this region results from the ejecta. The
abundances of the outer region are consistent with solar values
indicating that the X-ray emission is produced by swept-up
interstellar matter (ISM). The abundances obtained from the
spectra of the whole region show that the X-ray emission results
from a mixture of the ejecta and ISM. This remnant shows a
centrally peaked X-ray emission and extends to a radius of
$\sim$7.3 arcmin as seen in the XIS1 image (see Fig. 1). We will
discus the possible reasons for the centrally peaked emission in
subsection 5.2.

\subsection{Thermal emission}

From the thermal X-ray spectra of G272.2$-$3.2, we found a high
absorbing column density of $N_{\rm H}\sim 1.07\times10^{22}$$\rm
cm^{-2}$ that is in agrement with the value obtained by \citet
{b4} with {\it ROSAT} spectral fits and with the Galactic H\,{\sc
i} column density in that direction, $N_{\rm H}\sim
0.9\times10^{22}$$\rm cm^{-2}$ \citep {b17}. During the analysis
we let $N_{\rm H}$ vary for whole, center and outer parts to see
if there is a significant variation all over the remnant. We found
that the $N_{\rm H}$ value obtained for these three regions are
similar, indicating that there is not a significant density
gradient across the remnant. One reason for the high $N_{\rm H}$
value might be that the remnant's distance is large or there could
be molecular material or dust along the line of sight in that
direction, or both. However, in literature no such material has
been reported yet in that direction or in the vicinity of the
remnant. Considering these cases, we will use the upper limit
distance value, d=10 kpc, given by \citet {b4} throughout our
calculations.

The XIS spectra suggest an ionization time scale of $n_{\rm
e}t\sim6.5\times10^{10}$ $\rm cm^{-3}$s. For the full ionization
equilibrium, the ionization timescale, $\tau$=$n_{\rm e}t$, is
required to be $\geq$$10^{12}$ $\rm cm^{-3}$s \citep {b21}. The
value that we obtained for G272.2$-$3.2 shows that the plasma is
far from the full ionization equilibrium. We estimated the X-ray
emitting plasma volume of the remnant to be
$\sim$$1.2\times10^{60}f$ $\rm cm^{3}$, where $f$ is the volume
filling factor of the emitting gas within the SNR, we assumed the
emitting region to be a full sphere of radius 7.3 arcmin which is
our XIS spectral extraction region. Based on the emission measure
$EM=n_{\rm e}n_{\rm H}V$ determined by the spectral fitting, where
$n_{\rm e}$ and $n_{\rm H}$ are number densities of electrons and
protons respectively, and V is the X-ray emitting volume, and
assuming $n_{\rm e}=1.2n_{\rm H}$, we calculated the electron
density of the plasma $n_{\rm e}$ to be $\sim$0.48$f^{-1/2}$ $\rm
cm^{-3}$. The age of G272.2$-$3.2 calculated to be
$\sim$$4300f^{1/2}$ yr from t=$\tau$/$n_{\rm e}$. The mass of the
X-ray emitting plasma of G272.2$-$3.2 estimated to be $M_{\rm
x}=475f^{1/2}$ {M\sun} from equation $M_{\rm x}=n_{\rm e}V m_{\rm
H}$, where $m_{\rm H}$ is the mass of a hydrogen atom.

\subsection{Radial profile}

Centrally peaked X-ray morphology of G272.2$-$3.2 can be explained
by two models cloud evaporation \citep{b15} and thermal conduction
\citep{b16}. We first consider the cloud evaporation model of
\citet{b15}. According to this model, the SNR blast wave passes
over the cold clouds keeping them in the hot postshock gas. X-ray
emission arises from the gas evaporated from these shocked
 clouds. Second, we consider the radiative model of \citet{b16} also named as the ``fossil'' conduction
model. According to this model the hot plasma in the interiors of
the remnant gradually becomes uniform by thermal conduction and
detectable as centrally brightened in X-ray.

We studied radial variations of the electron temperature and
surface brightness profiles (see Fig. 3) to compare with these two
models. We see that there is no strong radial temperature
variation ($\sim$0.06 keV) and it is consistent with the
predictions of both evaporation and thermal conduction models.
Observed surface brightness variation which peaks at the center
($\sim$1.26$\times10^{-11}$ erg $\rm s^{-1}$$\rm cm^{-2}$$\rm
arcmin^{-2}$) and declines towards outer region
($\sim$0.38$\times10^{-11}$ erg $\rm s^{-1}$$\rm cm^{-2}$$\rm
arcmin^{-2}$) is consistent with the evaporation model. However,
in the vicinity of this remnant no molecular clouds or density
gradient of medium has been reported yet. This case is
inconsistent with the predictions of both models. The young age
($\sim$$4300f^{1/2}$ yr) of the remnant, in other words the NEI
condition of the plasma can not be explained by the thermal
conduction model which requires collisional ionization equilibrium
condition of the plasma. Therefore, considering all these, the
centrally peaked X-ray emission of this remnant may be explained
with the cloud evaporation model.

\subsection{Relative abundances in the ejecta}
Type Ia SN produces very small quantities of low-Z elements such
as Ne and Mg, and larger amounts of Si-group elements such as S
and Ca, and overabundant Fe and Ni as in our case with
G272.2$-$3.2. Therefore, we compare our best-fitting relative
abundances with the predicted nucleosynthesis yield of the
widely-used W7 and a delayed detonation (WDD2) Type Ia SN models
\citep{b11} as given in Fig. 4.

Although the abundances of Ne, Mg and Ca relative to Si are almost
consistent with both models, S and Ni relative to Si are higher
than the values that both models predict. The value of Fe relative
to Si is consistent with the WDD2 model, while it is much lower
than the value of W7 model predicts. The reason for the low value
of Fe might be that the entire Fe-rich core has not yet been
shocked, as is the case in SN 1006 \citep{b20}, Tycho \citep{b19}
and G337.2-0.7 \citep{b18}. Our results confirm that G272.2$-$3.2
has a Type Ia SN origin.

\section*{Acknowledgments}
AS is supported by the T\"{U}B\.{I}TAK PostDoctoral Fellowship.
This work is supported by the Akdeniz University Scientific
Research Project Management.

\onecolumn

\begin{figure}
\centering
  \vspace*{17pt}
  \includegraphics[width=12cm]{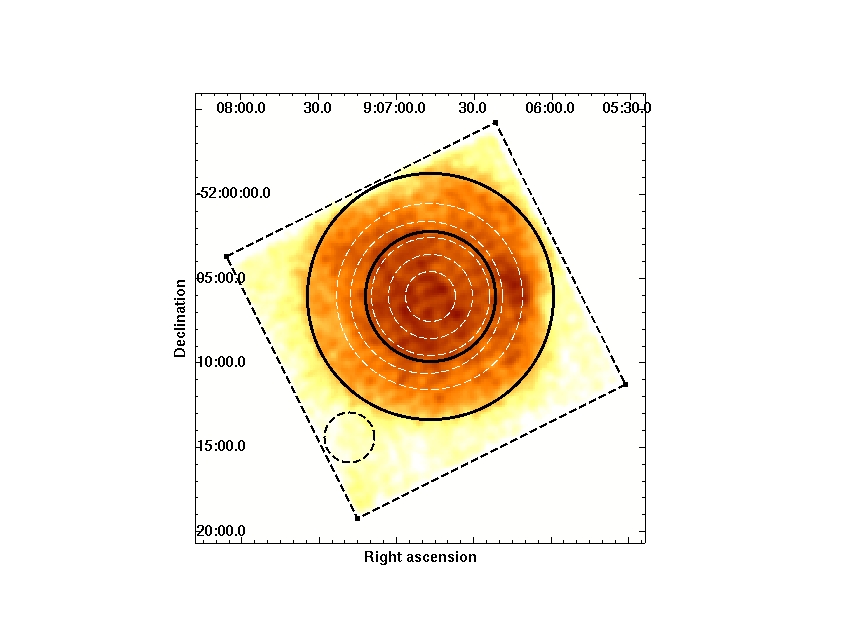}
  \caption{{\it Suzaku} XIS1 image of G272.2$-$3.2 in the 0.3$-$10 keV energy
  band. Solid black circles show central and outer regions, dashed white
  circles show the regions chosen to obtain the radial
variations of the electron temperature and surface brightness. The
black dashed circle represents the background region. The FOV of
the XIS1 is indicated by the black dashed
  square. The coordinates (RA and Dec.) are referred to epoch J2000.
   }
\end{figure}

\begin{figure}
\centering
  \vspace*{17pt}
  \includegraphics[width=8cm]{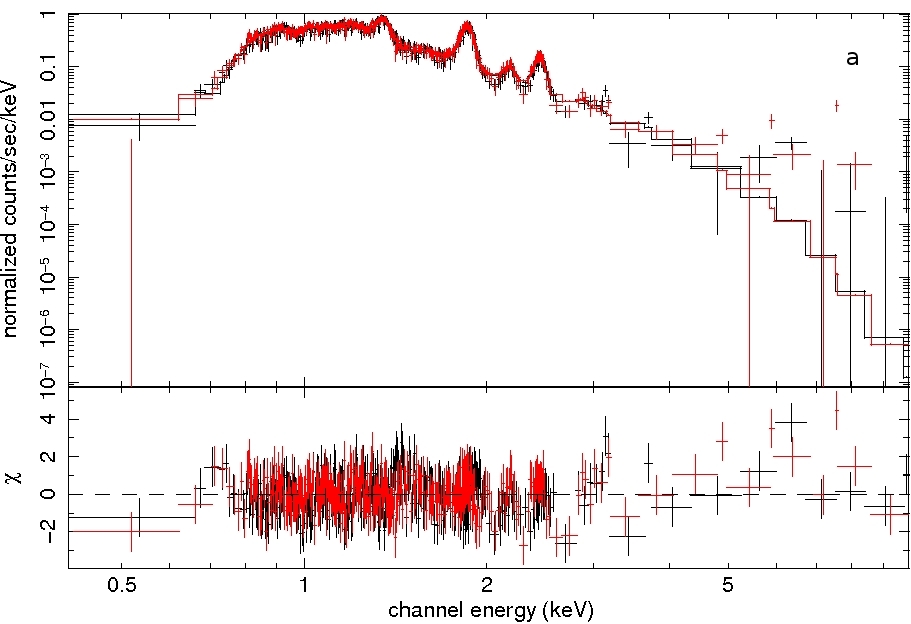}
  \includegraphics[width=8cm]{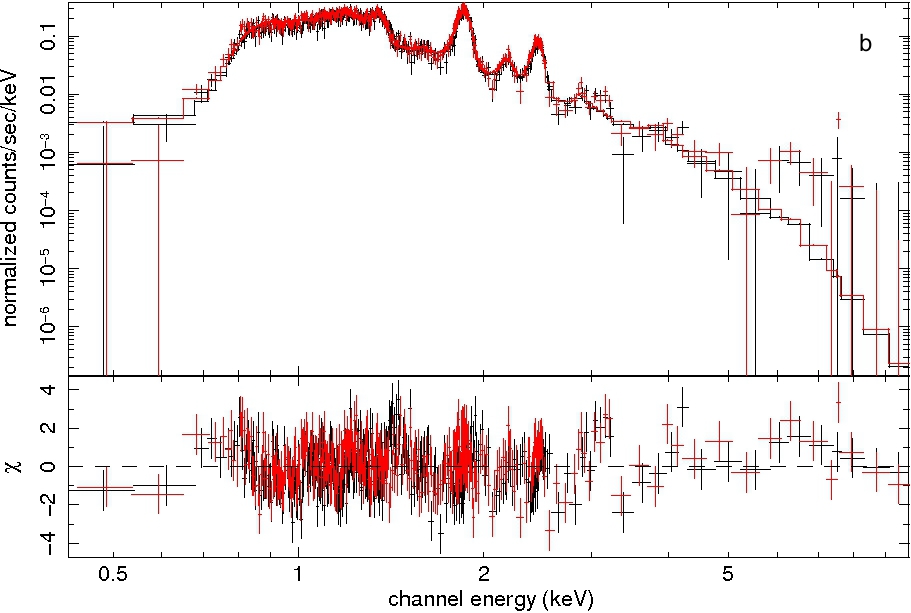}
  \includegraphics[width=8cm]{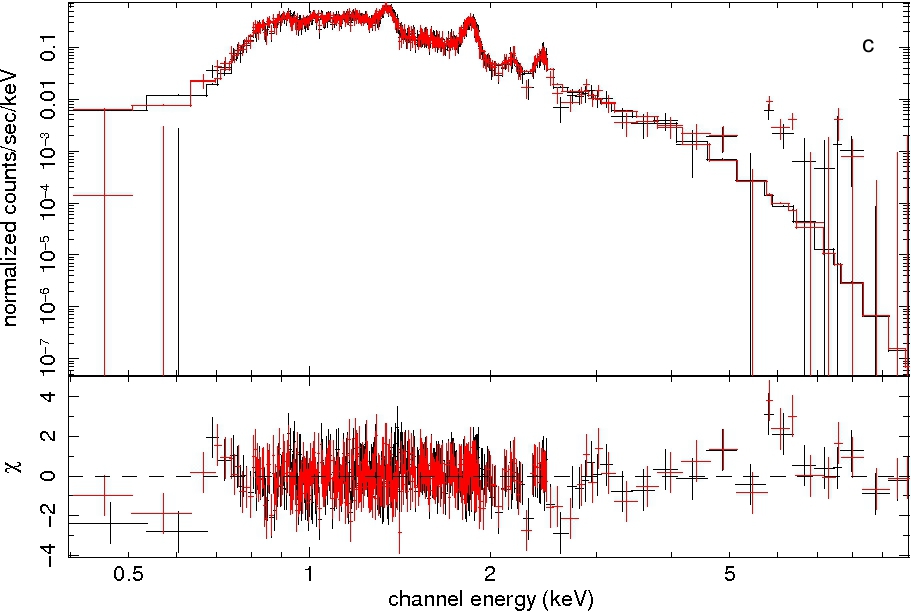}
  \caption{{\it Suzaku} FI (XIS0:red and XIS3:black) spectra of G272.2$-$3.2 in the 0.3$-$10 keV energy band.
  (a) Whole region, (b) Central region, (c) Outer region. The bottom windows give the residuals from the
best-fitting model for FI XIS spectra.}
\end{figure}

\begin{figure}
\centering
  \vspace*{17pt}
  \includegraphics[width=8cm]{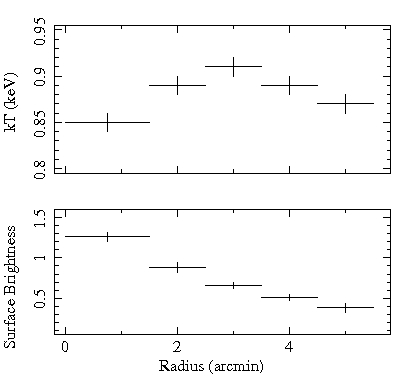}
  \caption{Radial variations of observed electron temperature and surface brightness of G272.2$-$3.2.
  Surface brightness is in the unit of ($\times10^{-11}$) erg $\rm s^{-1}$$\rm
cm^{-2}$$\rm arcmin^{-2}$.}
\end{figure}

\begin{figure}
\centering
  \vspace*{17pt}
  \includegraphics[width=8cm]{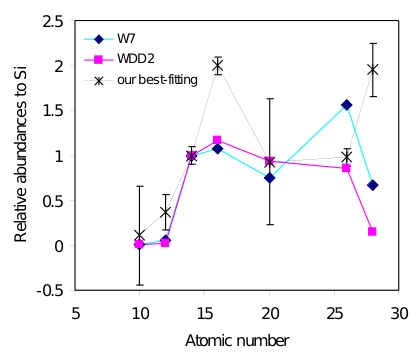}
  \caption{Best-fitting abundance ratios of Ne, Mg, Si, S, Ca, Fe and Ni relative to Si are shown by
  crosses and predicted abundance ratios from the carbon deflagration (W7,
\citet{b11}) model are shown by diamonds and from the delayed
detonation (WDD2, \citet{b11}) model are shown by squares.}
\end{figure}

\end{document}